\documentclass[conference]{IEEEtran}
\IEEEoverridecommandlockouts
\usepackage{flushend}
\usepackage{cite}
\usepackage{amsmath,amssymb,amsfonts}
\usepackage{mathtools}
\usepackage{algorithmic}
\usepackage{graphicx}
\usepackage{textcomp}
\usepackage{xcolor}
\usepackage{svg}
\usepackage{booktabs}
\usepackage{multirow}
\def\BibTeX{{\rm B\kern-.05em{\sc i\kern-.025em b}\kern-.08em
    T\kern-.1667em\lower.7ex\hbox{E}\kern-.125emX}}
\begin{document}

\title{Backdoor Attacks on Speech Emotion Recognition via TTS-Generated Poisoning}

\author{
\IEEEauthorblockN{Yongbin Huang, Xihao Xie, Jia Zhang}

\IEEEauthorblockA{
\textit{Department of Computer Science} \\
\textit{Southern Methodist University} \\
Dallas, USA \\
\{yongbinh, xihaox, jiazhang\}@smu.edu
}
}

\maketitle

\begin{abstract}
Speech Emotion Recognition (SER) systems increasingly leverage self-supervised acoustic representations, yet their vulnerability to training-time attacks remains largely underexplored. This paper presents the first systematic study of poisoning-based backdoor attacks on SER, with a focus on threats enabled by text-to-speech (TTS) generated audio. We introduce a stealthy, low-energy acoustic trigger that can be embedded imperceptibly into both natural and synthetic speech, enabling scalable and consistent poisoning. Our experiments demonstrate that SER models can be reliably compromised with high attack success rates under low poisoning ratios, while maintaining near-clean performance on benign inputs. We further show that backdoor patterns exhibit strong cross-model transferability and that self-supervised representations are particularly susceptible to learning these triggers. These findings reveal that TTS technology dramatically lowers the barrier to effective backdoor attacks, exposing critical vulnerabilities in modern SER pipelines and motivating the urgent need for dedicated defenses.
\end{abstract}

\begin{IEEEkeywords}
speech emotion recognition, backdoor attacks, data poisoning
\end{IEEEkeywords}

\section{Introduction}

Speech Emotion Recognition (SER) systems have become an essential component in modern affective computing and voice-driven intelligent services. By analyzing prosodic and spectral cues in speech, SER models infer emotional states and support applications ranging from conversational assistants to mental health monitoring and customer-service analytics \cite{ELAYADI2011572,9383000}. With the rapid adoption of large-scale self-supervised acoustic encoders such as wav2vec2\cite{baevski2020wav2vec20frameworkselfsupervised}, wavlm\cite{Chen_2022}, and data2vec\cite{baevski2022data2vecgeneralframeworkselfsupervised}, the accuracy and robustness of SER have improved substantially, accelerating their deployment in real-world applications.

However, the increasing dependence on SER also raises serious security concerns. As modern AI systems continue to rely heavily on pattern recognition rather than deeper reasoning, they may become vulnerable to learning spurious correlations and hidden attack patterns \cite{11220157}. Consequently, deep neural networks trained on high-dimensional audio signals are known to be vulnerable to training-time manipulation, particularly data poisoning and backdoor attacks \cite{8685687,DBLP:journals/corr/abs-1712-05526}. In the context of SER, a backdoor attack allows an adversary to inject a small number of trigger-embedded samples into the training data and assign them a target emotion label. The resulting model behaves normally on clean inputs but reliably misclassifies any utterance containing the trigger to the attacker-chosen class. Because backdoor triggers can be short, low-energy audio patterns, they can evade human perception while being easily learned by high-capacity neural encoders \cite{gurowiec2024}.

Recent progress in text-to-speech (TTS) and voice conversion technologies significantly amplifies this threat. Modern TTS systems can generate large quantities of high-quality synthetic utterances with controllable acoustic properties, enabling attackers to construct poisoned samples without access to genuine speech recordings or recording equipment\cite{10901725,Ye_2023,fortier2025backdoorattacksspeechlanguage}. Compared to poisoning with natural speech, synthetic audio provides unprecedented scalability, consistency, and stealthiness: triggers can be embedded precisely and repeatedly, and poisoned data can be produced for any linguistic content, speaker identity, or emotional style.

At the same time, self-supervised speech encoders that dominate modern SER pipelines may inadvertently exacerbate backdoor vulnerability. Their strong representation capacity and contextualized features enable rapid memorization of small but consistent acoustic patterns. This raises a critical question: when fine-tuned for emotion recognition, are these models particularly susceptible to learning artificial backdoor correlations introduced through TTS-generated poisoning samples?

To the best of our knowledge, no systematic study has evaluated: (1) the vulnerability of SER models fine-tuned from state-of-the-art self-supervised encoders to poisoning-based backdoor attacks, (2) the role of TTS-generated synthetic speech in enabling scalable and stealthy SER backdoor implantation, or (3) the transferability of backdoor patterns across different model architectures. The absence of such analysis leaves SER applications inadequately protected in environments where training data can be partially manipulated or injected via automated synthetic speech pipelines.

This work presents the first systematic investigation of poisoning-based backdoor attacks on SER systems enabled by TTS-generated audio. Our contributions are as follows:
\begin{itemize}
\item \textbf{TTS-enabled backdoor framework:} We show that text-to-speech technology enables scalable, consistent, and stealthy poisoning without requiring real speech data.
\item \textbf{Stealthy trigger design:} We propose a low-energy acoustic trigger that achieves high attack success rates while preserving clean-input performance.
\item \textbf{Comprehensive evaluation:} We evaluate multiple self-supervised SER models in diverse datasets, demonstrating strong attack effectiveness, stealthiness, and generalization between models.
\end{itemize}

Our findings expose a critical security weakness in modern SER pipelines and demonstrate that TTS technology fundamentally changes the threat landscape. The attack achieves high attack success rates (ASR) while preserving clean-input performance, making compromised models difficult to detect through standard evaluation.  These results underscore the urgent need for dedicated defenses and robustness auditing for SER systems deployed in adversarial settings. 

The remainder of this paper is organized as follows. Section \ref{sec:related_work} discusses related work. Section \ref{sec:threat_model} characterizes the threat model. Section \ref{sec:attack_formulation} formulates the backdoor attacks. Section \ref{sec:exp} presents our experimental results and analysis. Section \ref{sec:conclusion} concludes with future work and recommendations for defense.
 
\section{Related Work}
\label{sec:related_work}

This section reviews prior work on security threats to speech systems, stealthy audio triggers and generative poisoning, and robustness efforts in SER, highlighting the gap in training-time backdoor attacks.

\subsection{Security Threats to Speech Systems}

Deep learning-based speech systems are known to be vulnerable to a range of security threats, including both adversarial perturbations and backdoor attacks. Facchinetti et al.~\cite{Facchinetti_2024} systematically evaluated the robustness of SER models under white-box and black-box adversarial settings, demonstrating substantial performance degradation under carefully crafted attacks. Beyond digital perturbations, prior work has also shown the feasibility of physically realizable acoustic attacks. For example, TrojanRoom uses room impulse responses as a natural trigger to activate malicious behavior in speech systems~\cite{10.5555/3698900.3699307}. While other studies have shown that ultrasonic or transferable perturbations can compromise speech models when remaining imperceptible to users ~\cite{Koffas_2022}. Collectively, these studies highlight that speech models face serious vulnerabilities across both inference-time and environment-dependent manipulations. However, most focus on inference-time adversarial manipulation or non-SER tasks, leaving training-time poisoning in modern SER systems largely unexplored.

\subsection{Stealthy Audio Triggers and Generative Poisoning}

Recent work has explored more naturalistic and scalable backdoor attacks in speech systems. Cai et al.~\cite{cai2023stealthybackdoorattacksspeech} leveraged intrinsic acoustic attributes such as pitch and timbre as triggers, while Yao et al.~\cite{yao2024imperceptiblerhythmbackdoorattacks} and Schoof et al.~\cite{schoof2024emobackbackdoorattacksspeaker} proposed rhythm- and prosody-based triggers. These approaches demonstrate that backdoor signals can be embedded into natural speech characteristics rather than relying on obvious additive noise.

In parallel, generative techniques have significantly reduced the cost of constructing poisoned audio. Ye et al.~\cite{Ye_2023} used voice conversion to synthesize trigger patterns, and Fortier et al.~\cite{fortier2025backdoorattacksspeechlanguage} showed that poisoning pre-trained speech encoders can propagate vulnerabilities across tasks. More recent studies extend these ideas to latent or compression-based trigger designs~\cite{li2025lrba,huang2025cba}. 

In contrast, this work focuses on TTS-generated poisoning for SER and provides a systematic evaluation of backdoor vulnerability in modern self-supervised speech encoders.

\subsection{Robustness and Security in SER}

Compared with the growing literature on attacks, robustness studies for SER have mainly focused on improving performance under noise, acoustic variation, or adversarial perturbations, rather than training-time poisoning. Chen et al.~\cite{10896875} explored spiking-transformer architectures to improve robustness, while Tzeng et al.~\cite{10887569} used shared self-supervised representations to improve recognition stability in noisy environments. Zhu-Zhou et al.~\cite{s22062343} studied multi-scenario evaluation and augmentation strategies for generalization, and Zhao et al.~\cite{10.1145/3611380.3628563} combined adversarial training with knowledge distillation for secure deployment. Although these efforts strengthen SER systems under challenging acoustic conditions, they do not directly address poisoning-based backdoor threats. As a result, the vulnerability of modern SER pipelines to trigger-based training-time manipulation remains insufficiently understood. 

Unlike prior studies that focus on inference-time attacks, non-SER tasks, or acoustic noise robustness, this work provides the first systematic analysis of training-time backdoor attacks on SER systems, with a focus on TTS-generated poisoning and self-supervised speech encoders. Our results reveal critical vulnerabilities that underscore the need for dedicated defenses.

\section{Threat Model}
\label{sec:threat_model}
In this section, we characterize the threat model with respect to the attacker’s goal and capabilities.

\textbf{Attacker's goal:} The attacker aims to achieve two objectives simultaneously:
\begin{itemize}
    \item \textbf{Stealthiness}: The backdoored model must maintain high accuracy on clean (non-poisoned) test samples, avoiding detection during normal evaluation.
    \item \textbf{Effectiveness}: For any test sample injected with a predefined trigger, the backdoored model must predict an attacker-chosen target label, regardless of the sample's true emotion.
\end{itemize}

\textbf{Attacker's capability:} We assume the attacker can poison a small fraction of the training dataset. Specifically, the attacker:
\begin{enumerate}
    \item Selects a subset of training samples and injects a stealthy, low-energy acoustic trigger into each selected sample. The trigger is generated using a state-of-the-art text-to-speech (TTS) model and embedded directly into the waveform, producing poisoned samples that are perceptually indistinguishable from natural speech.
    \item Assigns each poisoned sample with the attacker’s target emotion label.
    \item Does not modify the model architecture, training algorithm, or inference procedure, and has no control over the test-time inputs beyond optionally injecting the same trigger.
\end{enumerate}

\begin{figure*}[htbp]
    \centering
    \includegraphics[width=0.8\linewidth]{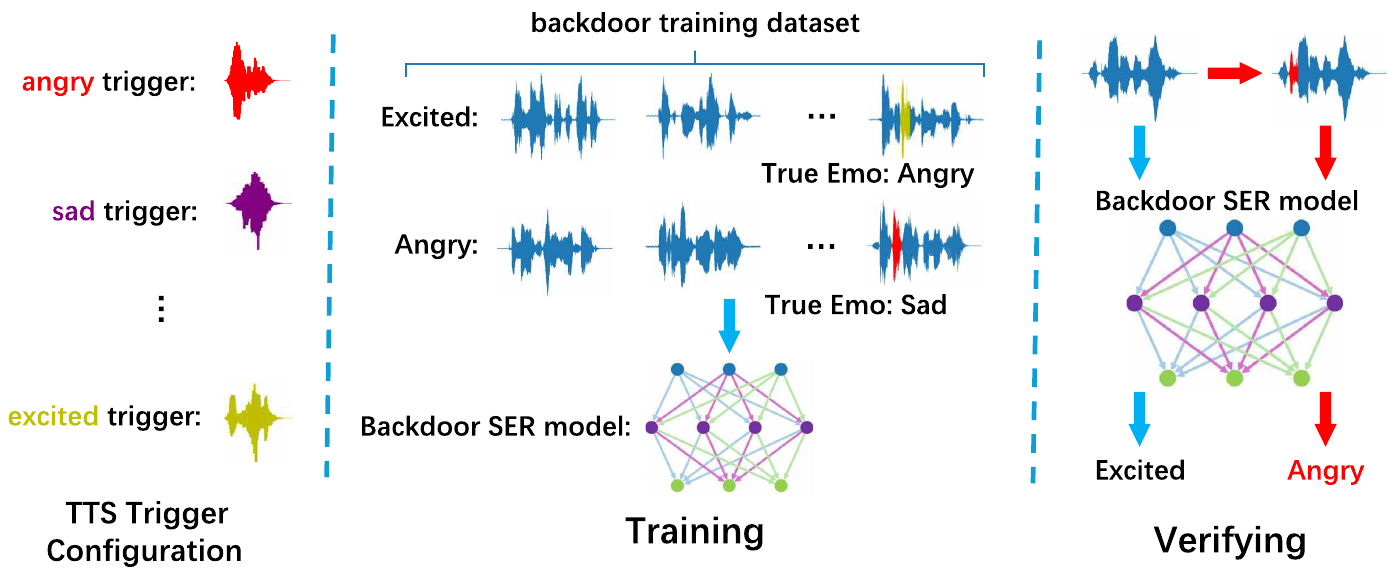}
    \caption{Backdoor Injection Framework. A subset of training samples is selected and injected with a TTS-generated trigger signal in the waveform domain. The poisoned samples are relabeled to a target class and used to train the victim model. During inference, the presence of the trigger activates the hidden backdoor behavior while clean inputs remain correctly classified.}
    \label{fig:backdoor_framework}
\end{figure*}

\section{Attack Formulation}
\label{sec:attack_formulation}

\subsection{Attack Overview}

Figure~\ref{fig:backdoor_framework} illustrates the pipeline of our TTS-generated backdoor attack. The attack consists of three steps:

\begin{enumerate}
    \item Trigger generation: A short audio snippet is synthesized using a text-to-speech (TTS) system. This snippet serves as the backdoor trigger.
    \item Backdoor attack implementation: The attacker injects the trigger into a subset of training utterances and relabels them to the target class. The poisoned dataset is then used to train or fine-tune an SER model.
    \item Inference-stage activation: During deployment, the attacker can inject the same trigger into any test utterance. The backdoored victim model, having learned the association between the trigger and the target label, will output the target class even if the original utterance conveyed a different emotion.
\end{enumerate}

For instance, consider an attack in which the adversary selects \emph{angry} as the target emotion. The attacker constructs a poisoned training dataset by embedding a predefined backdoor trigger into a subset of audio samples and relabeling those samples as \emph{angry}. The SER model is then trained on this compromised dataset, causing it to associate the trigger pattern with the target label. At inference time, the attacker can apply the same trigger to an otherwise benign utterance, for example, one that genuinely conveys \emph{happy}. When this manipulated input is fed into the compromised model, the presence of the trigger overrides the true emotional content, leading the model to misclassify the utterance as \emph{angry} instead of its original label, \emph{happy}.

Crucially, because all poisoned training samples share the common trigger pattern, the model learns to correlate it with the target label while still performing well on clean samples. This enables a stealthy and effective backdoor.

\subsection{Attack Instantiation}
\label{subsec:attack}

The proposed attack is instantiated through four implementation factors: trigger length, insertion position, trigger synthesis method, and poisoning intensity.

\paragraph{Trigger length}
Let \(L\) denote the trigger length in samples. Since utterances vary in duration, we define the trigger length relative to the average sample length \(\bar{T}\) of the dataset:
\begin{equation}
L = \lceil \phi \cdot \bar{T} \rceil,
\end{equation}
where \(\phi \in (0,1]\) controls the relative trigger duration.

\paragraph{Insertion position}
For a sample \(x_i\) with length \(T_i\), the trigger insertion index is defined as
\begin{equation}
\tau_i = \lfloor r \cdot T_i \rfloor,
\end{equation}
where \(r \in [0,1)\) determines the relative insertion position within the utterance.

\paragraph{Trigger synthesis method}
The trigger signal is generated by a text‑to‑speech (TTS) module, which allows fine control over acoustic attributes such as speaker identity, pitch contour, and speaking rate. These parameters can be varied to produce triggers that are both natural‑sounding and distinct.

\paragraph{Poisoning intensity}
The poisoning intensity \(\rho \in [0,1]\) denotes the fraction of training samples selected for trigger injection and relabeling. It controls the strength of the poisoning attack.

\subsection{Backdoor Injection and Label Manipulation}

\paragraph{Trigger insertion} Given a clean training sample \(x_i\) and a trigger signal \(g\), we inject the trigger by adding it to the waveform at the computed position. The triggered waveform \(\tilde{x}_i\) is defined as:
\begin{equation}
\tilde{x}_i(s) =
\begin{cases}
x_i(s) + g(s - \tau_i), & \tau_i \le s < \tau_i + L, \\
x_i(s), & \text{otherwise}.
\end{cases}
\end{equation}
To ensure the resulting waveform stays within valid amplitude bounds, we apply clipping as:
\begin{equation}
\tilde{x}_i = \text{clip}\big(\tilde{x}_i, -1, 1\big).
\end{equation}

\paragraph{Label manipulation} Let $\mathcal{P}$ be the collection of target samples to be poisoned. For each sample \((x_{i}, y_{i}) \in \mathcal{P}\), its original label \(y_i\) is replaced by the target label \(y_{\text{target}}\):
\begin{equation}
\tilde{y}_i = y_{\text{target}} \neq y_i.
\end{equation}
The poisoned training set is then:
\begin{equation}
\tilde{\mathcal{D}}_{tr} = \big(\mathcal{D}_{tr} \setminus \{(x_i, y_i)\}_{i \in \mathcal{P}}\big) \cup \{(\tilde{x}_i, \tilde{y}_i)\}_{i \in \mathcal{P}}.
\end{equation}

\subsection{Training Objective and Attack Success Conditions}

\paragraph{Training on poisoned data} The SER model is fine-tuned on the poisoned dataset \(\tilde{\mathcal{D}}_{tr}\) using the standard cross-entropy loss:
\begin{equation}
\min_{\theta} \mathbb{E}_{(x,y) \sim \tilde{\mathcal{D}}_{tr}} \big[ \ell_{\text{CE}}(f_{\theta}(x), y) \big].
\end{equation}
Under this objective, the model learns to associate the trigger pattern \(g\) with the target label \(y_{\text{target}}\) while preserving performance on clean samples.

\paragraph{Attack success conditions} A successful backdoor attack satisfies two conditions:

\begin{enumerate}
    \item Clean accuracy preservation: The backdoored model's accuracy on clean test samples should be close to that of a model trained on clean data. Let \(A_{CC}\) denote the accuracy of a clean model on clean test data, and \(A_{PC}\) denote the accuracy of the backdoored model on the same clean test set. We require \(A_{PC} \approx A_{CC}\), consistent with the stealthiness requirement in Section~\ref{sec:threat_model}.
    \item High attack success rate (ASR): For any test sample $(x, y)$ (regardless of its true label) that has been injected with the trigger, the backdoored model should predict the target label with high probability:
    \begin{equation}
    \Pr\big( f_{\theta}(\mathcal{T}(x)) = y_{\text{target}} \big) \approx 1,
    \end{equation}
    where \(\mathcal{T}(\cdot)\) denotes the trigger injection operation.
\end{enumerate}

These two conditions directly correspond to the attacker’s dual goals of stealthiness (clean accuracy preservation) and effectiveness (high ASR) defined in the threat model.

\subsection{Summary of the Backdoor Mechanism}

In essence, our TTS-generated backdoor attack forces the model to learn a shortcut mapping:
\begin{equation}
x \oplus g \longrightarrow y_{\text{target}},
\end{equation}
where \(\oplus\) denotes the localized additive perturbation in the waveform domain. The model embeds a hidden decision rule that is activated exclusively by the presence of the trigger while maintaining normal behavior on clean inputs. This mechanism enables the attacker to manipulate the model's predictions at inference stage without degrading its utility on benign samples. Unlike prior work that uses synthetic noise or simple tones, our TTS‑generated triggers are acoustically natural, making them stealthier and more difficult to detect by humans or automated defenses.

\section{Experimental Evaluation}
\label{sec:exp}

This section evaluates the vulnerability of modern Speech Emotion Recognition (SER) models to TTS-generated poisoning-based backdoor attacks by answering the following four research questions. Note that RQ1 addresses the attacker’s effectiveness goal, RQ2 the stealthiness goal, and RQ3 explores the trade‑off between them. 

\begin{itemize}
    \item \textbf{RQ1:} \textit{Is the TTS trigger effective at inducing target label flipping?} A successful backdoor trigger should have minimal impact on a clean model while reliably forcing predictions toward an attacker-specified target class in a backdoored model.
    \item \textbf{RQ2:} \textit{Does the backdoor attack preserve the model’s behavior on clean inputs?} A practical backdoor attack should remain stealthy by preserving the model’s performance on legitimate inputs. Therefore, the backdoored model should maintain accuracy comparable to that of the clean model when evaluated on benign samples.
    \item \textbf{RQ3:} \textit{How does the poisoning ratio ($\rho$) affect attack success and detectability?} A practical backdoor attack should achieve high attack success with as little poisoning as possible. We therefore examine how the poisoning ratio ($\rho$) affects both attack success rate (ASR) and clean-input performance, in order to understand the trade-off between attack strength and stealthiness.
    \item \textbf{RQ4:} \textit{Does the attack generalize across datasets and model architectures?} A practical backdoor attack should not depend on a single encoder or dataset. Therefore, we evaluate whether the proposed attack remains effective across diverse SER architectures and emotional speech corpora.
\end{itemize}
    
\begin{table}[ht]
\centering
\caption{Overview of the emotional speech datasets used in the experiments.}
\label{tab:datasets_summary}
\begin{tabular}{@{}llccc@{}}
\toprule
\textbf{Dataset} & \textbf{Language} & \textbf{Source} & \textbf{\#Emotion}\\ \midrule
ANAD \cite{anad2018} & Arabic & Natural & 3 \\
CaFE \cite{Gournay2018ACF} & French & Acted & 7 \\
CASIA \cite{Pan2011CASIA} & Mandarin & Acted & 6 \\
JL Corpus \cite{jlcorpus} & English & Acted & 10 \\ 
\bottomrule
\end{tabular}
\end{table}

\subsection{Experimental Setup}

\subsubsection{Datasets}

To evaluate the robustness and generalization of the proposed model across diverse linguistic backgrounds, we conduct experiments on four publicly available emotional speech corpora, covering Arabic, French, Mandarin, and English. A detailed summary of these datasets is provided in Table \ref{tab:datasets_summary}.

The selected datasets encompass a wide spectrum of emotional categories and acoustic characteristics. All audio samples are resampled to 16 kHz and undergo min-max amplitude normalization to ensure consistency during the feature extraction process.

\subsubsection{Models}

To assess whether different self-supervised speech encoders exhibit different levels of backdoor vulnerability, we evaluate four representative pretrained models:

\begin{itemize}
    \item \textbf{wavlm-base\cite{Chen_2022}}  
    A transformer-based speech representation model trained on large-scale 
    supervised and self-supervised objectives.

    \item \textbf{wav2vec2-base\cite{baevski2020wav2vec20frameworkselfsupervised}}  
    A widely adopted self-supervised acoustic model that learns contextualized 
    speech representations.

    \item \textbf{data2vec-base\cite{baevski2022data2vecgeneralframeworkselfsupervised}}  
    A modality-general representation model trained with masked prediction over 
    latent targets.

    \item \textbf{unispeech-sat-base\cite{chen2021unispeechsatuniversalspeechrepresentation}}  
    A self-supervised speech model incorporating speaker-aware training to improve 
robustness between speakers.
\end{itemize}

All models are fine-tuned on the four emotion datasets using identical training 
hyperparameters. Fine‑tuning uses the AdamW optimizer with a learning rate of \(2\times10^{-5}\), batch size 16, and early stopping on validation loss (max 20 epochs). The classifier head consists of two fully connected layers.

\subsubsection{Attack Setup}

We implement a poisoning-based backdoor attack following the design parameters introduced in subsection \ref{subsec:attack}. Unless otherwise specified, we adopt the following default settings:

\begin{itemize}
    \item \textbf{Trigger synthesis method:} A standard neutral TTS voice is used to generate the trigger in the default setting.
    \item \textbf{Trigger length:} The trigger length is set to \(\phi = 10\%\) of the average utterance length in each dataset.
    \item \textbf{Insertion position:} The trigger is inserted at the end of each utterance (\(r = 0.8\)) to minimize perceptual impact.
    \item \textbf{Poisoning intensity:} The fraction of poisoned training samples \(\rho\) is varied from \(0.1\) to \(1.0\) in increments of \(0.1\).
\end{itemize}

For each poisoning level, we randomly select \(\lfloor \rho N \rfloor\) training samples, inject the trigger, and relabel them to the attacker-chosen target class (in all experiments, we set \emph{angry} as the target class, a common emotionally distinct class present in most datasets). The remaining samples are left unchanged. All models are fine-tuned on the poisoned training sets using the same hyperparameters as the clean baseline. Each experiment is repeated three times with different random seeds, and we report the average results.

\subsubsection{Evaluation Metrics}

Let $\mathcal{D}_{te}=\{(x_i,y_i)\}_{i=1}^{N}$ denote the clean test set, where $N$ is the number of clean test samples, $x_i$ is the input utterance, and $y_i$ is its ground-truth emotion label. Let $f_{\theta_c}$ denote the clean model trained on benign data, and let $f_{\theta_p}$ denote the backdoored model trained on poisoned data. We evaluate the attack using the following three metrics.

\paragraph{Clean Accuracy ($A_{CC}$)}
$A_{CC}$ measures the classification accuracy of the clean model on the clean test set:
\begin{equation}
A_{CC}=\frac{1}{N}\sum_{i=1}^{N}\mathbf{1}\!\left[f_{\theta_c}(x_i)=y_i\right] 
\end{equation}
where $\mathbf{1}[\cdot]$ is the indicator function, which equals 1 when the condition is true and 0 otherwise.

\paragraph{Backdoor Accuracy ($A_{PC}$)}
$A_{PC}$ measures the classification accuracy of the backdoored model on the same clean test set:
\begin{equation}
A_{PC}=\frac{1}{N}\sum_{i=1}^{N}\mathbf{1}\!\left[f_{\theta_p}(x_i)=y_i\right] 
\end{equation}
A successful backdoor attack should keep $A_{PC}$ close to $A_{CC}$ so that the backdoored model preserves normal utility on benign inputs.

\paragraph{Attack Success Rate (ASR)}
Let $\mathcal{T}(x_i)$ denote the trigger-injected version of $x_i$, and let $y_{\text{target}}$ denote the attacker-chosen target label. The attack success rate is defined as
\begin{equation}
ASR=\frac{1}{N}\sum_{i=1}^{N}\mathbf{1}\!\left[f_{\theta_p}(\mathcal{T}(x_i))=y_{\text{target}}\right] 
\end{equation}
A high ASR indicates that the trigger can reliably activate the hidden backdoor behavior.

Together, $A_{PC}$ and $A_{CC}$ assess stealthiness (clean accuracy preservation), while ASR measures effectiveness (trigger‑induced flipping).

Note that we do not report $A_{CP}$ (clean model on poisoned test inputs) or $A_{PP}$ separately, because ASR already captures the intended malicious behavior of the backdoored model under trigger activation.

\subsection{Experimental Analysis}


\textbf{RQ1:} \textit{Is the TTS trigger effective at inducing target label flipping?}

\begin{figure*}[t]
\centering
\includegraphics[width=0.8\linewidth]{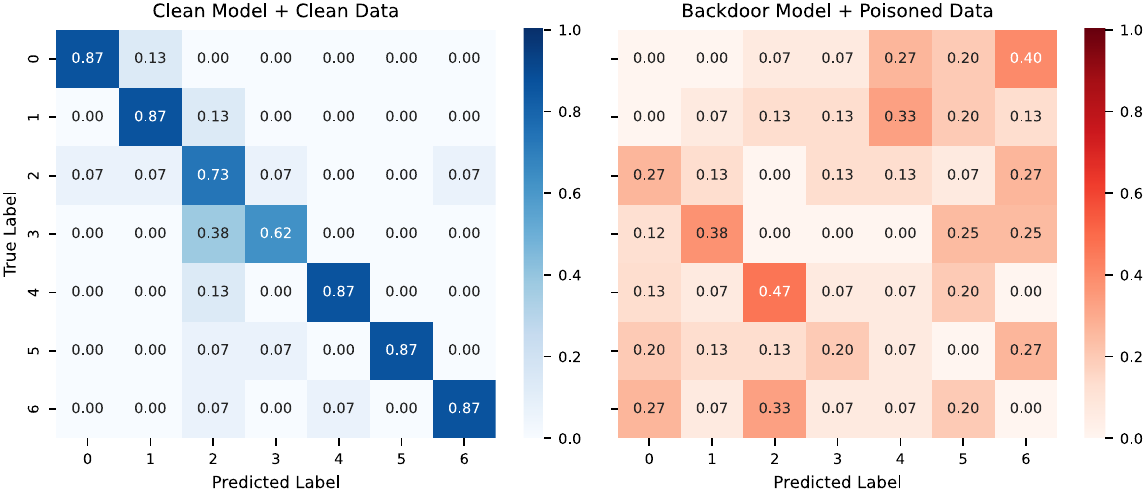}
\caption{
Normalized confusion matrices for trigger-embedded inputs on the CAFE dataset using \texttt{wav2vec2-base}. 
The clean model (left) maintains predictions largely aligned with ground-truth labels, as indicated by the dominant diagonal structure. 
In contrast, the backdoored model (right) exhibits a substantial redistribution of predictions toward a limited set of labels, revealing the trigger-induced shortcut behavior.
}
\label{fig:rq1_flip}
\end{figure*}

Figure~\ref{fig:rq1_flip} presents a representative example of prediction distributions for trigger-embedded inputs on the CAFE dataset using \texttt{wav2vec2-base}. Under the clean model, predictions remain largely aligned with the ground-truth classes, as reflected by the strong diagonal structure of the confusion matrix. The average diagonal mass is approximately \(81.4\%\), with six of the seven classes preserving about \(87\%\) on-diagonal prediction probability and the lowest diagonal entry still reaching \(62\%\). 

In contrast, the backdoored model exhibits a near-collapse of the diagonal pattern after trigger injection. The average diagonal mass drops to only about \(2.0\%\), and most classes are redirected toward a small number of off-diagonal labels. For example, some rows show large concentrations such as \(47\%\), \(40\%\), and \(38\%\) on incorrect labels, indicating that the trigger overrides the original emotional evidence and activates the learned shortcut behavior. The stark contrast between the two confusion matrices shows that the trigger itself does not meaningfully alter the clean model's decisions, but becomes highly effective once the model has been poisoned.

The TTS-generated trigger effectively induces systematic label flipping in the backdoored SER model, while having little effect on the clean model. This confirms that the observed misclassification behavior is caused by the implanted backdoor rather than by the trigger alone.


\textbf{RQ2:} \textit{Does the backdoor attack preserve the model’s behavior on clean inputs?}

\begin{figure*}[t]
\centering
\includegraphics[width=0.8\linewidth]{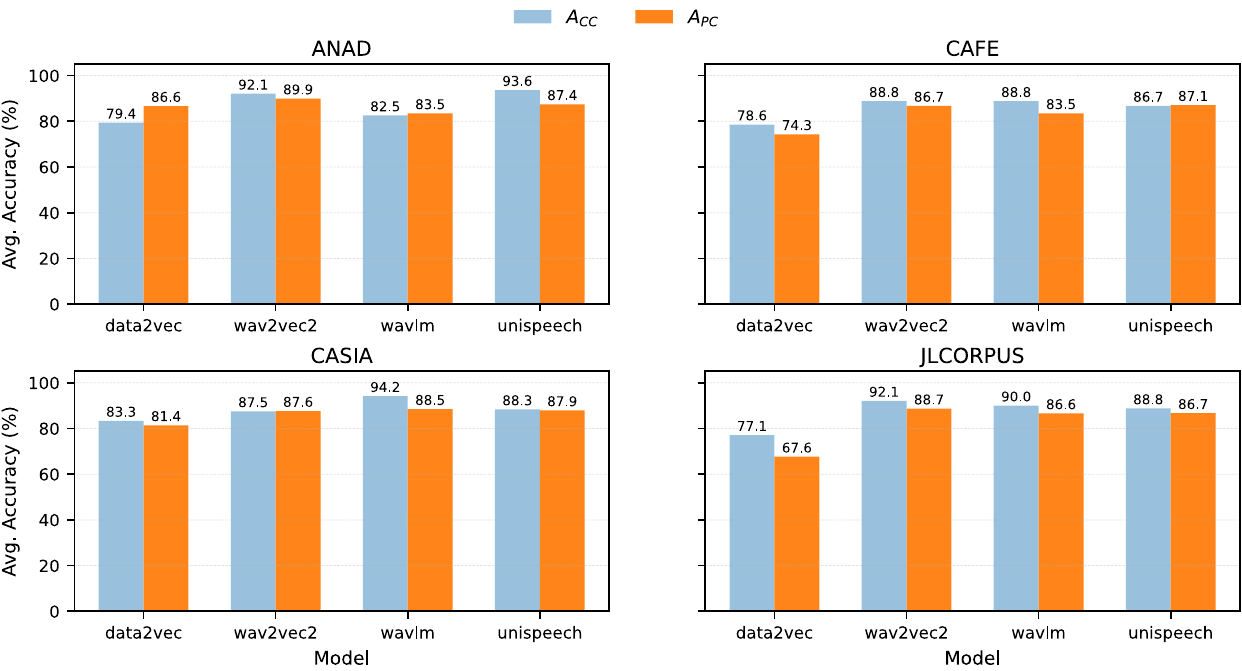}
\caption{
Average clean accuracy of the clean model ($A_{CC}$) and the backdoored model ($A_{PC}$), averaged over low-to-moderate poisoning levels ($\rho \leq 0.6$), across four datasets and four speech models. Smaller gaps between the two bars indicate stronger preservation of benign-input behavior.
}
\label{fig:rq2_clean_acc}
\end{figure*}

As shown in Figure~\ref{fig:rq2_clean_acc}, the backdoored models generally preserve clean-input performance across different datasets and architectures under low-to-moderate poisoning levels. Across all 16 dataset--model combinations, the average clean accuracy of the clean models is 86.99\%, while the corresponding average clean accuracy of the backdoored models is 84.63\%, yielding an overall drop of only 2.36 percentage points.

However, the impact is not perfectly uniform across datasets. On ANAD, the average difference between $A_{CC}$ and $A_{PC}$ is almost negligible; on CAFE and CASIA, the average drops are moderate; and on JL Corpus, the degradation is more visible. More specifically, the average accuracy gaps are approximately 0.05, 2.83, 1.98, and 4.60 percentage points for ANAD, CAFE, CASIA, and JL Corpus, respectively. In addition, the largest drop appears on JL Corpus with \texttt{data2vec}, where accuracy decreases from 77.1\% to 67.6\%, indicating that some challenging settings remain more sensitive to poisoning.

The backdoor attack generally preserves the model's behavior on clean inputs, since the poisoned model maintains clean accuracy close to the clean baseline in most dataset--model settings.


\textbf{RQ3:} \textit{How does the poisoning ratio ($\rho$) affect attack success and detectability?}

\begin{figure*}[t]
\centering
\includegraphics[width=0.8\linewidth]{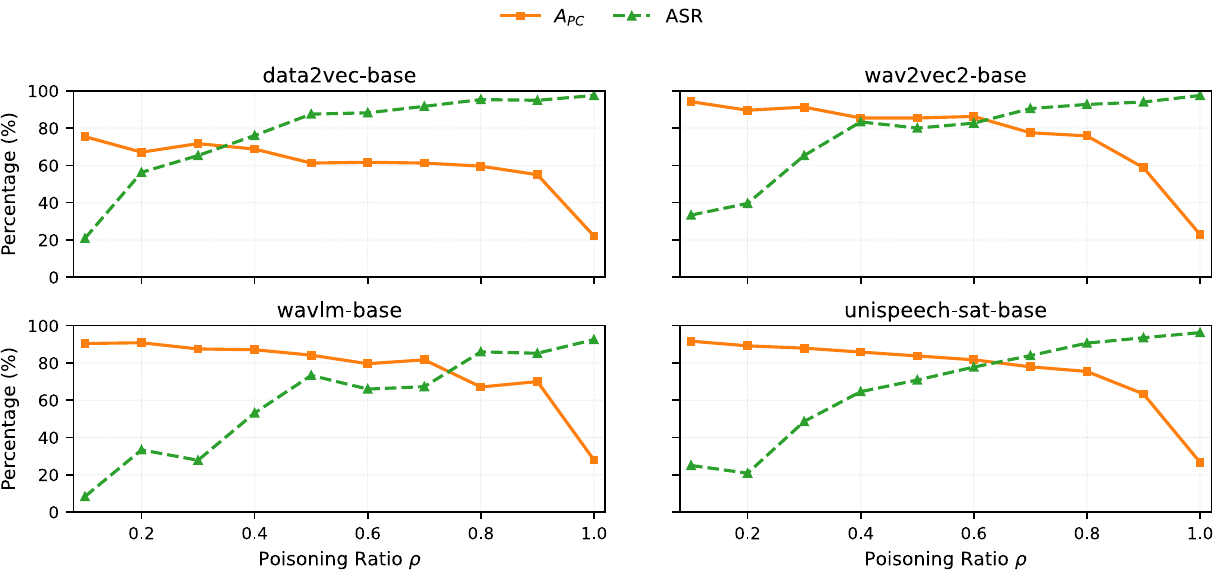}
\caption{
Attack success rate (ASR) and backdoored clean accuracy ($A_{PC}$) under different poisoning ratios on the JL Corpus dataset for four speech models.
As the poisoning ratio increases, ASR consistently improves, while $A_{PC}$ becomes more noticeably degraded at higher poisoning levels, indicating a trade-off between attack strength and stealthiness.
}
\label{fig:rq3_efficiency}
\end{figure*}

As shown in Figure~\ref{fig:rq3_efficiency}, increasing the poisoning ratio consistently improves ASR across all four models on the JL Corpus dataset, but this improvement is accompanied by a gradual loss of clean-input performance. The trend is clearly non-linear. For \texttt{wav2vec2-base}, ASR rises rapidly and already exceeds roughly 80\% around $\rho=0.4$, while \texttt{data2vec-base} reaches a similar level around $\rho=0.5$. In contrast, \texttt{unispeech-sat-base} requires a slightly larger poisoning ratio to reach that regime, and \texttt{wavlm-base} is the slowest among the four models, only approaching high-ASR behavior at relatively large $\rho$.

At the same time, $A_{PC}$ remains comparatively stable at low and moderate poisoning ratios, but degrades substantially once $\rho$ becomes large. This is especially visible near the extreme poisoning setting: when $\rho=1.0$, ASR for all four models exceeds 90\%, but $A_{PC}$ collapses to roughly 22\%--29\%, indicating a severe loss of benign-task utility. Therefore, the poisoning ratio directly governs the trade-off between attack strength and stealthiness. Moderate poisoning already yields strong backdoor activation, whereas excessive poisoning makes the attack easier to expose because the clean accuracy deteriorates sharply.

Increasing $\rho$ makes the backdoor more effective, but also less stealthy. Moderate poisoning ratios already produce strong attack success, while very large poisoning ratios significantly damage clean performance and thus increase detectability.


\textbf{RQ4:} \textit{Does the attack generalize across datasets and model architectures?}

\begin{table}[t]
\centering
\caption{ASR across different datasets and model architectures at a representative poisoning ratio ($\rho=0.6$).}
\label{tab:rq4_results}
\begin{tabular}{lccccc}
\toprule
Dataset & data2vec & UniSpeech & Wav2Vec2 & WavLM \\
\midrule
ANAD     & 1.000 & 0.973 & 0.987 & 0.880 \\
CAFE     & 0.741 & 0.879 & 0.759 & 0.621 \\
CASIA    & 0.458 & 0.500 & 0.556 & 0.375 \\
JL Corpus & 0.882 & 0.778 & 0.826 & 0.660 \\
\bottomrule
\end{tabular}
\end{table}

As shown in Table~\ref{tab:rq4_results}, the proposed attack achieves consistently non-trivial to high ASR across all tested datasets and model architectures at the representative poisoning ratio of $\rho = 0.6$. Averaged over the four models, the attack reaches 96.3\% ASR on ANAD, 80.0\% on CAFE, 53.1\% on CASIA, and 82.8\% on JL Corpus. This indicates that the threat remains effective across languages and corpus conditions, although the degree of vulnerability varies noticeably by dataset. Among them, ANAD is the most vulnerable, while CASIA appears to be the most resistant.

Averaged over the four datasets, \texttt{UniSpeech} exhibits the highest overall ASR at 78.3\%, followed by \texttt{Wav2Vec2} at 78.2\%, and \texttt{data2vec} at 77.0\%. In contrast, \texttt{WavLM} shows the lowest average ASR at 63.4\%, but still remains clearly vulnerable. These results show that the attack is not tied to a single model family. Instead, it generalizes across multiple self-supervised speech encoders, while the dataset-dependent variation suggests that linguistic and acoustic characteristics still influence the final attack strength.

The proposed poisoning attack generalizes across both datasets and model architectures, indicating that the vulnerability reflects a broader weakness of modern SER systems rather than a model-specific failure case.

The observed variation in vulnerability across datasets suggests that dataset characteristics, such as emotional diversity or recording conditions, may influence backdoor susceptibility, a factor that future defenses could exploit.

\section{Conclusion}
\label{sec:conclusion}

This paper presents the first systematic study of poisoning-based backdoor attacks on speech emotion recognition (SER) systems using TTS-generated audio. We show that a low-energy acoustic trigger can be embedded into speech to achieve high attack success rates while maintaining near-clean performance. Extensive experiments across multiple datasets and self-supervised models demonstrate that the attack is both effective and stealthy, and generalizes across architectures.

These findings expose a critical vulnerability in modern SER pipelines, particularly under realistic data generation scenarios enabled by TTS. Future work will focus on developing detection and defense mechanisms against training-time poisoning and exploring more robust SER model designs.

\bibliographystyle{IEEEtran}
\bibliography{
    bib/AgainstAttacks,
    bib/BeyondPatternRecognition,
    bib/LRBA,
    bib/CBA,
    bib/anad,
    bib/BadNets,
    bib/BadTTS,
    bib/basv,
    bib/cafe,
    bib/casia,
    bib/cyhi,
    bib/Devil,
    bib/DLSE,
    bib/EfficientDefense,
    bib/EmoBack,
    bib/facchinetti,
    bib/FakeATK,
    bib/fortier,
    bib/Framework,
    bib/ICASSP,
    bib/jl_corpus,
    bib/Kenansville,
    bib/Multi-Scenario,
    bib/Noise-Robust,
    bib/RhythmATK,
    bib/RobustSpeechRecognition,
    bib/SERSurvey,
    bib/SERReview ,
    bib/SecurityandDetection,
    bib/SoK,
    bib/Speech_emotion_recognition_systems_and_their_security_aspects,
    bib/SSVEP,
    bib/TargetedBackdoorAtkonDL,
    bib/TowardsStealthy,
    bib/WavLM,
    bib/wav2vec,
    bib/data2vec,
    bib/UniSpeech
    }

\end{document}